%% file: mnras_template.tex
\documentclass[fleqn,usenatbib]{mnras}

\usepackage{newtxtext,newtxmath}

\usepackage[T1]{fontenc}

\DeclareRobustCommand{\VAN}[3]{#2}
\let\VANthebibliography\thebibliography
\def\thebibliography{\DeclareRobustCommand{\VAN}[3]{##3}\VANthebibliography}

\usepackage{graphicx,xcolor,verbatim} 
\usepackage{amsmath}	

\input{affiliations}

\title{JWST Observations of Starbursts: Cold Clouds and Plumes Launching in the M\,82 Outflow}

\author[Fisher et al. ]{Deanne B. Fisher$^{1,2}$, Alberto D. Bolatto$^{3,4}$, John Chisholm$^5$, Drummond Fielding$^{6,7}$, Rebecca C. Levy$^{8}$\thanks{\AAPF}, \newauthor Elizabeth Tarantino$^{9}$, Martha L. Boyer$^{9}$, Serena A. Cronin$^3$, Laura A. Lopez$^{10,11}$, J.D. Smith$^{12}$,\newauthor Danielle A. Berg$^5$, Sebastian Lopez$^{10,11}$, Sylvain Veilleux$^{3,4}$, Paul P.\ van der Werf$^{13}$, Torsten B\"oker$^{14}$, \newauthor Leindert~A.~Boogaard$^{15}$, Laura Lenki\'{c}$^{16}$, Simon C.~O.\ Glover$^{17}$, Vicente Villanueva$^{18}$, Divakara Mayya$^{19}$,  \newauthor Thomas S.-Y. Lai$^{20}$, Daniel~A.~Dale$^{21}$, Kimberly L. Emig$^{28,29}$, Fabian Walter$^{15}$, Monica Rela\~no$^{23}$, \newauthor Ilse De Looze$^{24}$, Elisabeth A. C. Mills$^{25}$, Adam K. Leroy$^{10}$, David S. Meier$^{26,27}$, Rodrigo Herrera-Camus$^{18}$, \newauthor Ralf S.\ Klessen$^{17,30}$
\\
$^1$ \Swinburne \\
$^2$ \ASTROTD \\
$^3$\Maryland \\
$^4$ \JSI \\
$^5$ \UTA \\
$^6$\Flatiron \\
$^7$\Cornell \\
$^8$\Arizona \\
$^{9}$\STScI \\
$^{10}$\OSU \\
$^{11}$\OSUCCAPP\\
$^{12}$\UToledo\\
$^{13}$\Leiden\\
$^{14}$\ESOST\\
$^{15}$\MPIA \\
$^{16}$\JPL\\
$^{17}$\ITA \\
$^{18}$\UdeC \\
$^{19}$\INAOE \\
$^{20}$\IPAC \\
$^{21}$\UWYO \\
$^{23}$ \UGR \\
$^{24}$ \UGent\\
$^{25}$ \Kansas \\
$^{26}$ \NMMT \\
$^{27}$ \NRAOSocorro \\
$^{28}$ \France \\ 
$^{29}$ \ParisObs \\
$^{30}$ \IWR
}

\date{Accepted XXX. Received YYY; in original form ZZZ}

\pubyear{2024}

\begin{document}
\label{firstpage}
\pagerange{\pageref{firstpage}--\pageref{lastpage}}
\maketitle

\begin{abstract}
In this paper we study the filamentary substructure of 3.3~$\mu$m PAH emission from JWST/NIRCam observations in the base of the M\,82 star-burst driven wind. We identify plume-like substructure within the PAH emission with widths of $\sim$50~pc.  Several of those plumes extend to the edge of the field-of-view, and thus are at least 200--300~pc in length. In this region of the outflow, the vast majority ($\sim$70\%) of PAH emission is associated with the plumes. We show that those structures contain smaller scale ``clouds" with widths that are $\sim$5--15~pc, and they are morphologically similar to the results of ``cloud-crushing" simulations. We estimate the cloud-crushing time-scales of $\sim$0.5--3~Myr, depending on assumptions. We show this time scale is consistent with a picture in which these observed PAH clouds survived break-out from the disk rather than being destroyed by the hot wind. The PAH emission in both the midplane and the outflow is shown to tightly correlate with that of Pa~$\alpha$ emission (from HST/NICMOS data), at the scale of both plumes and clouds, though the ratio of PAH-to-Pa~$\alpha$ increases at further distances from the midplane. Finally, we show that the outflow PAH emission is suppressed in regions of the M\,82 wind that are bright in X-ray emission. Overall, our results are broadly consistent with a picture in which cold gas in galactic outflows is launched via hierarchically structured plumes, and those small scale clouds are more likely to survive the wind environment when collected into the larger plume structure.  
\end{abstract}






\section{Introduction} \label{sec:intro}
Large scale galactic winds contribute to the enrichment of the circumgalactic medium and play critical roles in regulating star formation by removing gas from the star-forming midplane of galaxies \citep{Chevalier1985, Heckman1990, Veilleux2005,Tumlinson2017,Rupke2018,Veilleux2020}. This process is frequently invoked in models of galaxy evolution as a necessary component to match basic properties of galaxies, such as the stellar mass function and star formation rates in galaxies across cosmic history \citep{hopkins2012model,Pillepich2018}. Indeed, it is a consensus view in modern theories of galaxy evolution that winds driven by stellar feedback are a required component \citep[for a review see][]{Naab2017}. 

Stellar feedback driven winds are launched from small-scale regions within the disk and are observed to extend many kiloparsecs above the plane. The wind is composed of a hot ($10^6$~K), highly ionized gas \citep{Strickland2009} which is surrounded by a cone of warm ($10^4$~K) ionized material \citep{Shopbell1998}, and further out by cold gas  \citep{Walter2002,Bolatto2013Natur,Leroy2015,Martini2018}. The complexity of galactic winds lies in both the wide range of phases and the large dynamic range in spatial scales that play roles in shaping the kinematics and evolution of the gas \citep[e.g.][]{Fielding2022}. Multiphase, resolved studies of outflows are direly needed but are observationally challenging. 

An extra layer of complexity in understanding winds comes from observations over the past decade \citep{Chisholm2015,Heckman2015, McQuinn2019, Marasco2023} that consistently show that very little of the gas in winds escapes from the virial radius of the halo. It then follows that gas in wind re-accretes onto the disk in a so-called ``galactic fountain" scenario \citep{Shapiro1976,Fraternali2017}. Winds therefore play a regulatory role, keeping the gas out of the disk for a period of time before it later becomes available for future star formation. The timescales for this process are not well-known. Characterising the energetics, evolution, and substructure of the galactic winds in detail is thus critical, as these properties will determine the amount of time  for which the gas is outside of the galaxy as well as how the redistribution of baryons will ultimately shape galaxy growth.

The cold phase of galactic winds is the dominant mass component of gas that exists above the plane of the disk \citep{Veilleux2020}. \cite{bolatto2013} used early ALMA observations to make resolved observations of the wind in NGC\,253. They found gas connected to the hot X-ray wind and extends $\sim$300-500~pc, with a significant mass-loading of $\dot{M}_{\rm out}/SFR\sim10-15$. Similar results were found for M\,82 \citep{Leroy2015,Martini2018}. \cite{Krieger2021} used high resolution ($\sim$1~arcsec) observations of M\,82 to find CO-bright clouds of gas extending over a kiloparsec from the midplane of the galaxy. 

Theoretically explaining how the hot winds accelerate cold gas sufficiently to survive breakout from the disk, without destroying the cold clouds, has historically been challenging \citep{Klein1994,Scannapieco2015,Schneider2017,Gronnow2018}. Yet recent work in idealised ``cloud-crushing" or ``wind-tunnel" simulations has shown that the addition of cooling in the physics of the gas can play a critical role in cloud survival \citep{gronke2018,Armillotta2017}. In these simulations a cold (T$<10^4$~K) over-density of gas is placed in a hot (T$\sim10^6$~K) wind fluid.  Subsequent theoretical work argues that cloud size is a key parameter for the survivability of cold gas \citep{gronke2020,Sparre2020,Abruzzo2022,Farber2022}. However, these results are heavily dependent on the physics used in each simulation. Recent semi-analytic theory argues that the relationship between the cold clouds and hot phases may be critical to determining the energetic properties and evolution of winds \citep{Fielding2022}. It is therefore necessary to determine if the substructure of cold gas predicted in simulations is comparable to that of observed galactic winds. However, observing such clouds directly requires very high spatial resolution on the cold phase gas. 

M\,82 is arguably the best-studied galactic wind in the literature. This is due to both its proximity ($D= 3.6$~Mpc) and its orientation on the sky. M\,82 is very near to edge-on (inclination ${\sim} 80^{\circ}$), which has allowed for the two nebulae extending outward from the stellar midplane of the galaxy to be easily observed \citep{Lynds1963}. It has, therefore, been well studied in cold \citep[e.g.,][]{Walter2002,Engelbracht2006,Chisholm2016,Martini2018,Krieger2021,Levy2023}, warm \cite[e.g.,][]{Shopbell1998} and hot gas phases \citep[e.g.,][]{Strickland2009,Lopez2020}. 

In this work, we will focus on new observations from the James Webb Space Telescope (JWST) of the polycyclic aromatic hydrocarbon (PAH) emission at the center of M\,82. The data set is presented in \citet{Bolatto2024arXiv}. PAH emission traces relatively cold gas (${\sim}10^3$~K) compared to the hot and warm-hot wind fluid in outflows. \cite{Engelbracht2006} showed, using observations from the \textit{Spitzer Space Telescope}, that the PAHs in M\,82 extend well outside the galaxy and are bright in the superwind. The PAH emission has filamentary sub-structure, which \cite{Leroy2015} show is in good agreement with the CO emission at large radii. This suggests that PAHs may be a good tracer of cold gas in outflows. High spatial resolution observations of cold gas in outflows have historically been difficult, as observations of CO or HI require interferometers. These observations are hampered by the ability to recover diffuse emission and surface brightness sensitivity. The availability of PAH and dust emission traced by JWST opens a new window into observing the cold component of outflows at high spatial resolution ($\sim$1--10~pc in the nearest starburst galaxies). 

In this paper we adopt a distance of 3.6~Mpc, such that 1~arcsec corresponds to 17.5~pc.

\section{Methods}

\subsection{JWST NIRCam Images}
\begin{figure*}
\begin{center}
\includegraphics[width=0.48\textwidth]{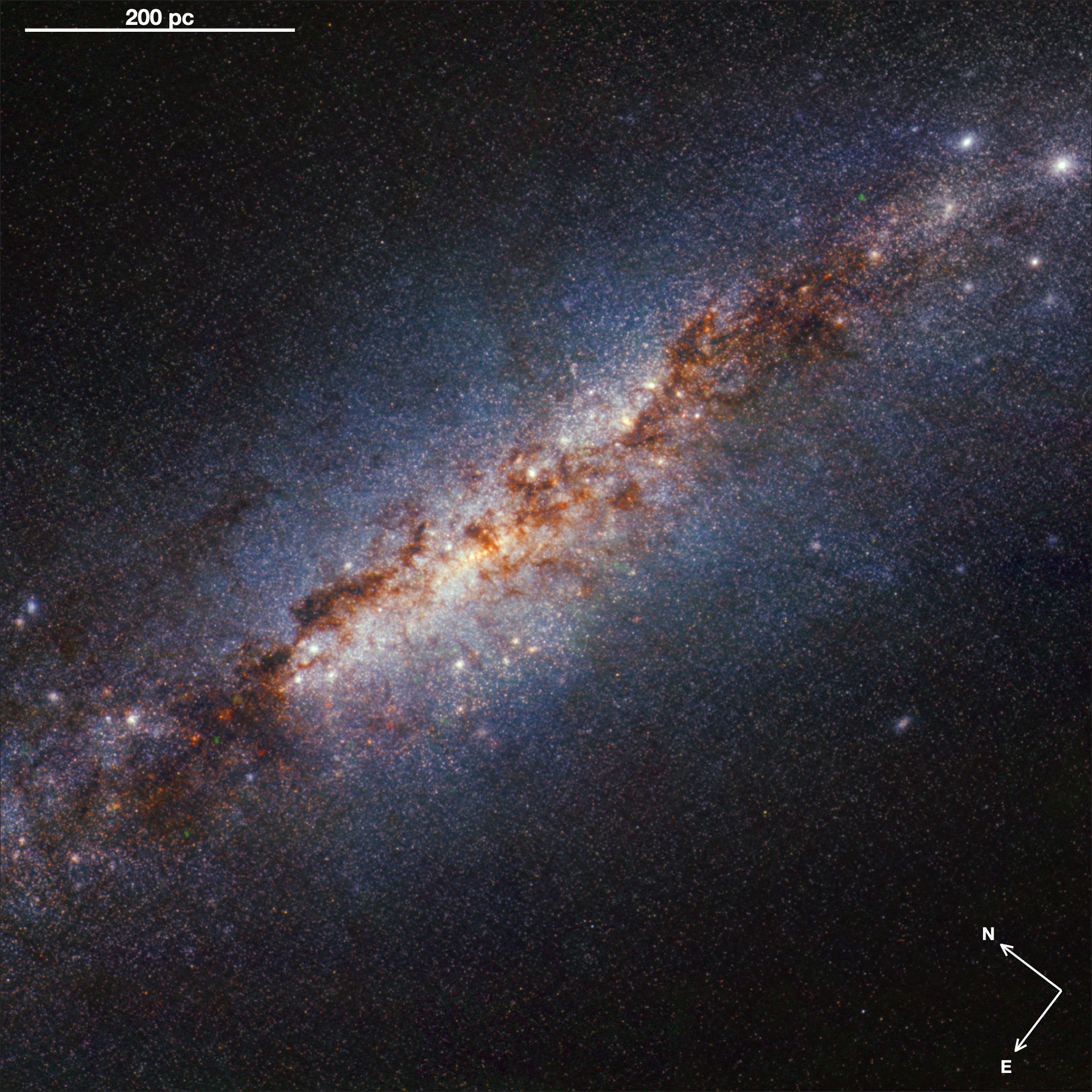}
\includegraphics[width=0.48\textwidth]{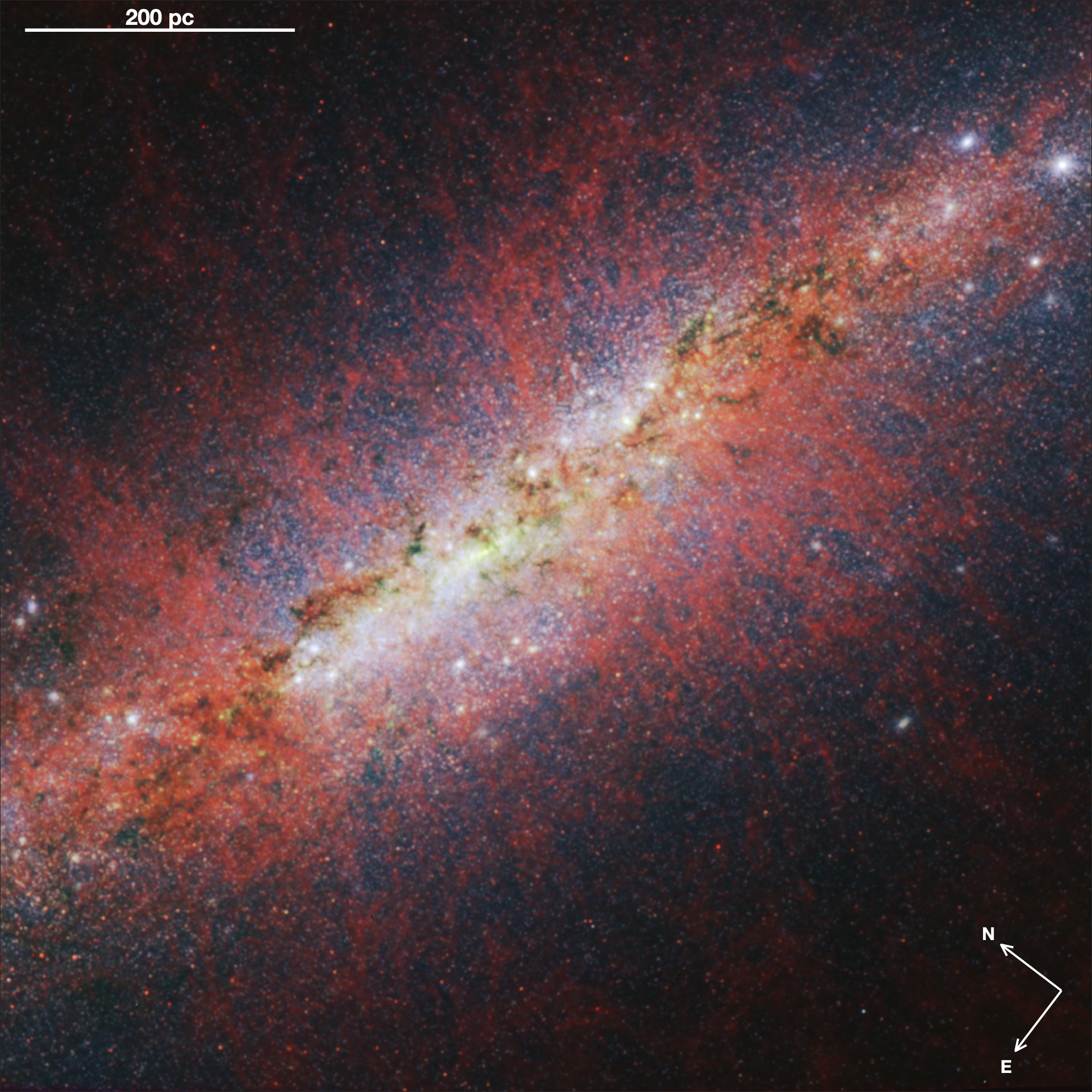}
\end{center}
\caption{M\,82 NIRCam observations, reproduced from \citet{Bolatto2024arXiv}. The left panel shows a combination of F212N, F164N, and F140M for red, green, and blue respectively. The right panel shows F335M, F250M, and F164N for red, green, and blue respectively. The most significant difference in the two images comes from F335M filter, which is dominated by the PAH emission. There are clear filaments of PAH extending out of the mid-plane of the galaxy and connecting to the larger scale wind. \label{fig:pretty} }
\end{figure*}

In this paper we focus on the NIRCam F335M image as a tracer of PAH emission in the center of M\,82. We also use F250M and F360M images to quantify the continuum level within the F335M filter. These dithered images have short exposures to avoid saturation in the bright central region of M\,82. The observations are part of the JWST Cycle 1 GO program \#1701 (PI: Bolatto), which images the outflows in M\,82 and NGC\,253 \citep[see][]{Bolatto2024}. The paper describes reduction of the images in more detail, and is summarised here for completeness. 

The NIRCam data used in this paper were taken in October 2022, and cover only the center of M\,82\footnote{Automatic pipeline processed MAST mosaics can be found under \url{http://dx.doi.org/10.17909/cwtn-nh63}}. We used a single SUB640 subarray observation with the RAPID readout, 4 intramodulebox primary dithers and 4 small-grid-dither subpixel dithers with 6 groups/integration.

The F335M targets the PAH feature at 3.3~$\mu$m. An advantage of analyzing the shortest wavelength PAH feature at 3.3 $\mu$m is that it can be observed with higher spatial resolution than other transitions observed with at longer wavelength. For M\,82, we can  use the F335M flux as a means of tracing the morphology of the PAH emission, and thus the colder (${\sim}10^2-10^3$~K) gas in the outflow, at parsec scales.  \cite{Draine2021} show that the $3.3$~$\mu$m feature is very sensitive to smaller grain size and lower ionization states. The 3.3~$\mu$m flux can be affected by ionization and grain size, and the conversion to PAH mass is not well known in winds. We, therefore, do not use 3.3~$\mu$m flux alone as a dust mass tracer in this work.      
It is unclear how 3.3~$\mu$m flux will behave in the unique environment at the base of the outflow, outside the disk but still near to the starburst.  Future work by our team will investigate the PAH flux ratios in the outflows of M\,82 and NGC\,253. 

We processed the NIRCam uncalibrated data products with the JWST pipeline version 1.9.6 \citep{Bushouse2023} and CRDS context jwst\_1077.pmap.  We employed all the default parameters for Stage 2 processing. We used the JWST/HST Alignment Tool \citep[JHAT;][]{Rest2023} to align the individual exposures. First, we aligned the F250M exposures to the stars available in Gaia-DR3 using JHAT. Then, we created a catalog of the stars from the F250M image by selecting pixels with brightness between $1$ and $250$\,MJy\,sr$^{-1}$. These stars are used to align the other frames. We applied the Stage 3 step in the pipeline to the aligned files with the {\tt tweakreg} step turned off, which yielded the final mosaics used in this work. We convolved the F335M and F250M images to match the point-spread-function (PSF) of the F360M. This results in a PSF of our PAH image of 0.118~arcsec. 


Extracting the PAH flux from the F335M image requires removal of underlying continuum. Ideally, this is carried out using continuum bands that bracket the emission feature. In our case, we have F250M and F360M images. However, \cite{Lai2020} show that the broad wing of the PAH emission feature can be significant in star-forming environments, and partially falls in the F360M bandpass. As described in \cite{Bolatto2024}, we follow a similar procedure as others \citep[e.g.][]{Sandstrom2023} in which we scale emission from the F335M filter to remove PAH and aliphatic contributions from the F360M filter.

It is very difficult to estimate a variance in the continuum subtracted image, as there are no regions without flux. We measure the standard deviation in several small (radius 0.1~arcsec) regions near the edge of the field-of-view that appear to not have significant sub-structure. We find values of $\sim$0.2-0.4~MJy~sr$^{-1}$, which we can take as an upper-bound estimate on the flux density uncertainty   in the image. We note this is very similar to the expected uncertainty of $\sim$0.25~MJy~sr$^{-1}$ from the JWST ETC. 

\subsection{NICMOS F187N}
To measure ionized hydrogen emission we use the HST NICMOS F187N and F160W for continuum subtraction. The data were taken as part of PID:7919 (PI: Sparks), and are described in \cite{Boker1999}\footnote{\url{http://dx.doi.org/10.17909/jdx7-qg88}}. The NIC3 observations cover a sufficient field of view to roughly match the NIRCam observations in this paper. NIC3 has lower angular resolution than our JWST data, with FWHM of order $\sim$0.3~arcsec. 

We carry out continuum subtraction using the F160W continuum filter. We follow a similar procedure as outlined in \cite{Boker1999}, in which the flux in F160W is scaled to the flux of F187N. The flux in F187N will be a combination of Paschen~$\alpha$ emission and stellar continuum, and we therefore must find an appropriate scaling factor to subtract the continuum. For a given F160W flux there is a well-defined minimum of the F187N flux. The pixels with values near the minimum F187N/F160W are assumed to indicate the scaling between the continuum. The pixels with larger F187N/F160W are those more dominated by Paschen~$\alpha$ emission. We fit a linear correlation between F187N and F160W (F187N$_{\rm cont}$ = $a \times$F160W + $b$)  for only those pixels in the lowest 0.5\% of F187N/F160W. We ensure that this captures a range of flux values, and then subtract F187N$_{\rm cont}$ from the F187N image.

Image registration with the NICMOS data requires an extra step, due to an offset in the WCS of the NICMOS data compared to the JWST NIRCam data. There were an insufficient number of point sources spread across the image for automated detection methods, as the mid-plane of the galaxy is blended at NICMOS resolution. We therefore used a by-eye match of point sources in the image in order to shift the WCS of the NICMOS continuum image to minimize the offset to the F250M image. We found the shift to be $\sim$0.96~arcsec in RA and $\sim$0.5~arcsec in DEC. We then applied the same offsets to the F187N image. We note that there were not enough point sources at large distance perpendicular to the mid-plane of M\,82 to carry out any higher order corrections to match the images. However, the distortion corrections in NICMOS multidrizzle are very well characterised and we assume this correction, after using standard pipeline methods, is minimal. 

\subsection{NOEMA CO(1--0)}
We use the CO(1--0) moment 0 map reported in \cite{Krieger2021}, and refer readers to the original publication for more details. \cite{Krieger2021} produced a map that covers $\sim$7~arcmin on the major-axis and $\pm$2.8~arcmin on the minor axis. This area is much larger than that of our current JWST data, but will be well-matched to our upcoming wide-scale observations of M\,82 with NIRCam and MIRI. 
The resulting CO(1--0) data cube has a median root-mean-square noise in a 5~km~s$^{-1}$ channel of 138~mK (or 5.15~mJy~beam$^{-1}$) and spatial resolution of $\sim$1.9~arcsec. 

\subsection{Chandra X-Ray Data}
We use data from \cite{Lopez2020} to characterize the X-ray emission in M\,82, and refer to their paper for further details on the observations and data reduction. In short, \cite{Lopez2020} combine ACIS-S observations into an equivalent exposure time of 544~ks producing a broad-band X-ray image (0.5-7.0~keV). The X-ray images are point sources are removed. The FWHM of the on-axis PSF of Chandra is $\sim$0.492~arcsec. As discussed by the authors, the resulting map shows a diffuse halo that is in broad agreement with the generally known morphology of the M\,82 wind. In our work, we will only use their map to study the location of X-ray peaks in relation to the cold gas tracer PAH. 

\section{PAH Plumes Extending From M\,82 Midplane}

\begin{figure*}
\begin{center}
\includegraphics[width=\textwidth]{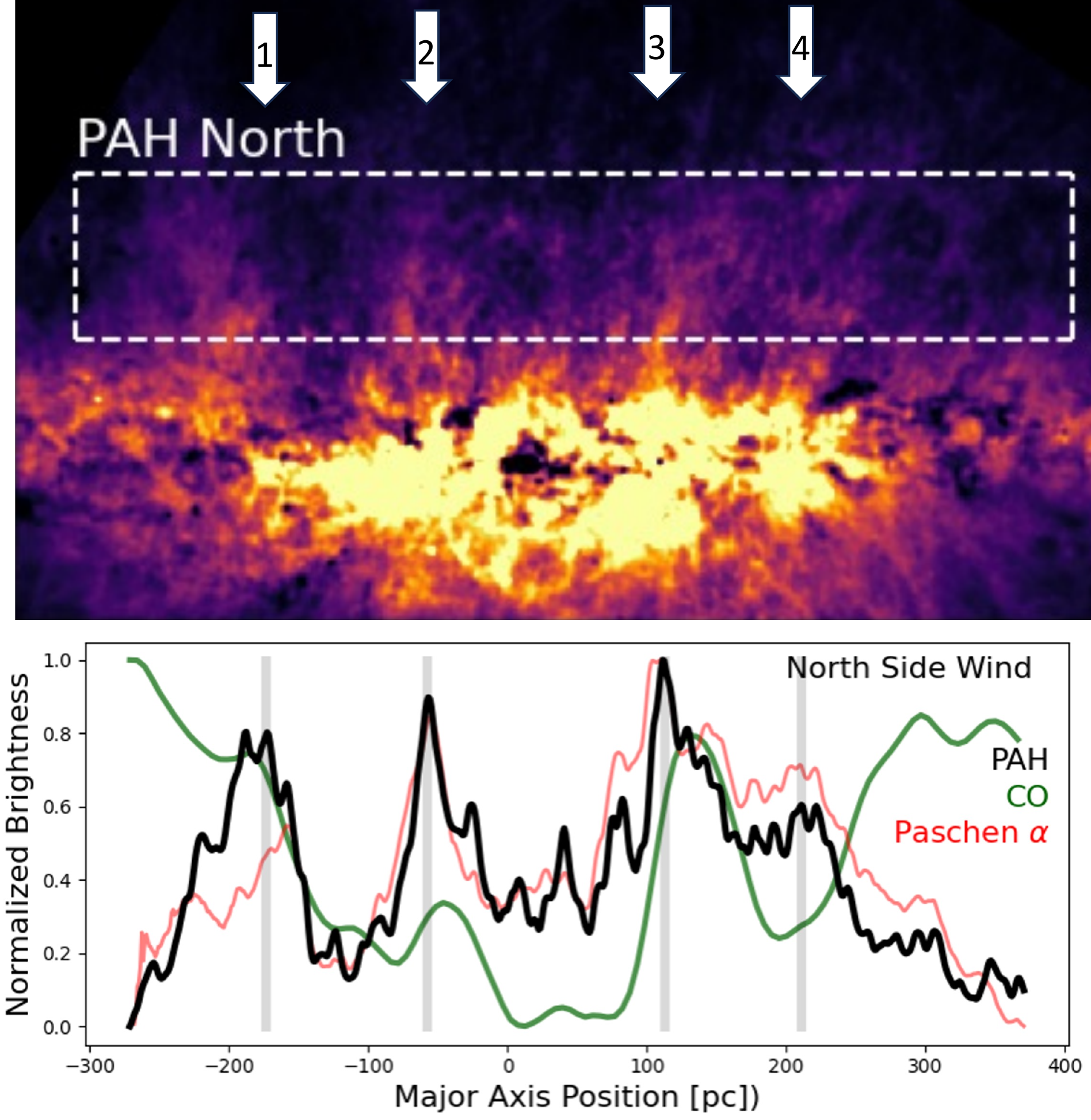}
\end{center}
\caption{The {\bf top panel} shows the PAH image. The white dashed rectangle represents the region analysed in the bottom panel of this figure. The {\bf bottom panel} shows horizontal slices of normalised flux for PAH (black), Pa~$\alpha$ (red) and CO moment 0 (green). We normalize the flux such that the highest value in the slice is unity and the lowest value in the slice is zero. The aim of these diagrams is to identify the location of peaks in PAH. These are shown as vertical grey lines and numbered arrows in top panel. \label{fig:northslice} }
\end{figure*}

\begin{figure*}
\begin{center}
\includegraphics[width=\textwidth]{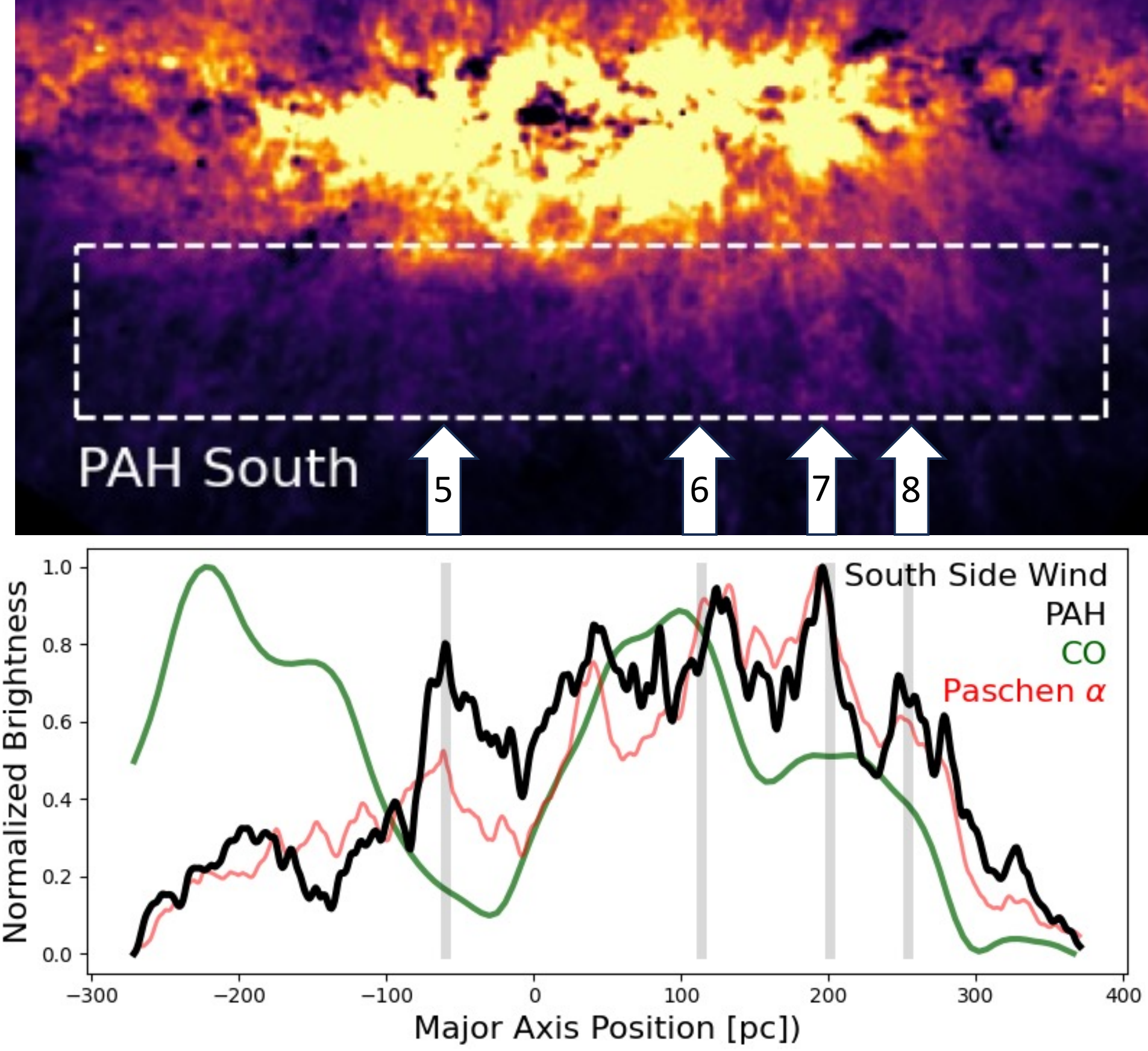}
\end{center}
\caption{ Same as Fig.~\ref{fig:northslice} for southern side of M\,82.    \label{fig:southslice} }
\end{figure*}

In Fig.~\ref{fig:pretty} we reproduce three-color images of M\,82 from NIRCam observations. \citet{Bolatto2024} present the overall NIRCam data set, and include a qualitative discussion of the filamentary substructure in M\,82, which we will study in more detail in this work. Using our assumed distance of 3.6~Mpc the spatial resolution of the F335M is of order $\sim$1.8~pc.  PAH emission has been observed to distances of $\pm$6~kpc from the disk \citep{Engelbracht2006,McCormick2013}. Our image in Fig.~\ref{fig:pretty}, therefore, represents the inner 25\% of the cold outflow. 

Overall, we identify a significant amount of filamentary gas plumes that are elongated away from the disk. We define a ``plume" as a coherent peak in the flux extending roughly perpendicular from the plane of the disk. \citet{Bolatto2024} identify several large (${\sim} 100$~pc) over-densities of PAH emission. These are similar to the structures identified by \cite{Wills1999} in their VLA radio continuum observations, which is likewise shown with updated radio continuum in \citet{Bolatto2024}.  On the north side of the image (upper half in Fig.~\ref{fig:pretty}~and~\ref{fig:northslice}) the 4 peaks in the normalized brightness slice are associated to plumes that are spaced at intervals of $\sim$100-150 from each other and extend outward to the edge of the field. These plumes extend and connect to the PAH emission structures observed at larger distances \citep{Engelbracht2006,McCormick2013}. 

The southern outflow of M\,82 also has filamentary structure. It is not as equally spaced. We note there are known differences in the mass loading and substructure of northern and southern outflows \citep{Walter2002,Leroy2015, Chisholm2016}, and they have for many years been known to be asymmetric \citep{Shopbell1998}.

In Figs.~\ref{fig:northslice}~\&~\ref{fig:southslice} we identify the locations of PAH emission peaks as arrows in the image and grey stripes in the panels below. The panels show horizontal slices of flux of the PAH emission that are centered $\sim$135$\pm50$~pc above the midplane (defined by stars). This height is chosen to be beyond the 50\% height of the star-light, but still having good coverage within the image. The flux is normalised such that the peak value in each slice is unity and the minimum value in each slice is zero. The intention is to determine if the locations of the plumes in Fig.~\ref{fig:pretty} are consistent with true peaks in the flux. 
We find in the normalized brightness profiles that in the region between z=80~pc and 180~pc (shown as boxes in Figs.~\ref{fig:northslice}~\&~\ref{fig:southslice}) the typical flux inside the plumes is 2-4~times brighter than in the areas between the plumes.

The location of the peaks in PAH are typically near to peaks in both CO and Pa~$\alpha$. \cite{Bolatto2024arXiv} qualitatively discuss the morphological similarity. There is a much tighter relationship  3.3~$\mu$m PAH with Pa~$\alpha$ than PAH and CO. We expand on PAH-Pa~$\alpha$ relationship more in \S5 of this paper.  The 17-times more coarse spatial resolution of the CO data, and the relatively small field of view makes any morphological comparison difficult. In Fig.~\ref{fig:northslice} there are local peaks in CO emission at horizontal positions of -180, -45 and 133~pc. These are all within 10~pc of the center of PAH plumes.


\begin{figure}
    \centering
    \includegraphics[width=0.45\textwidth]{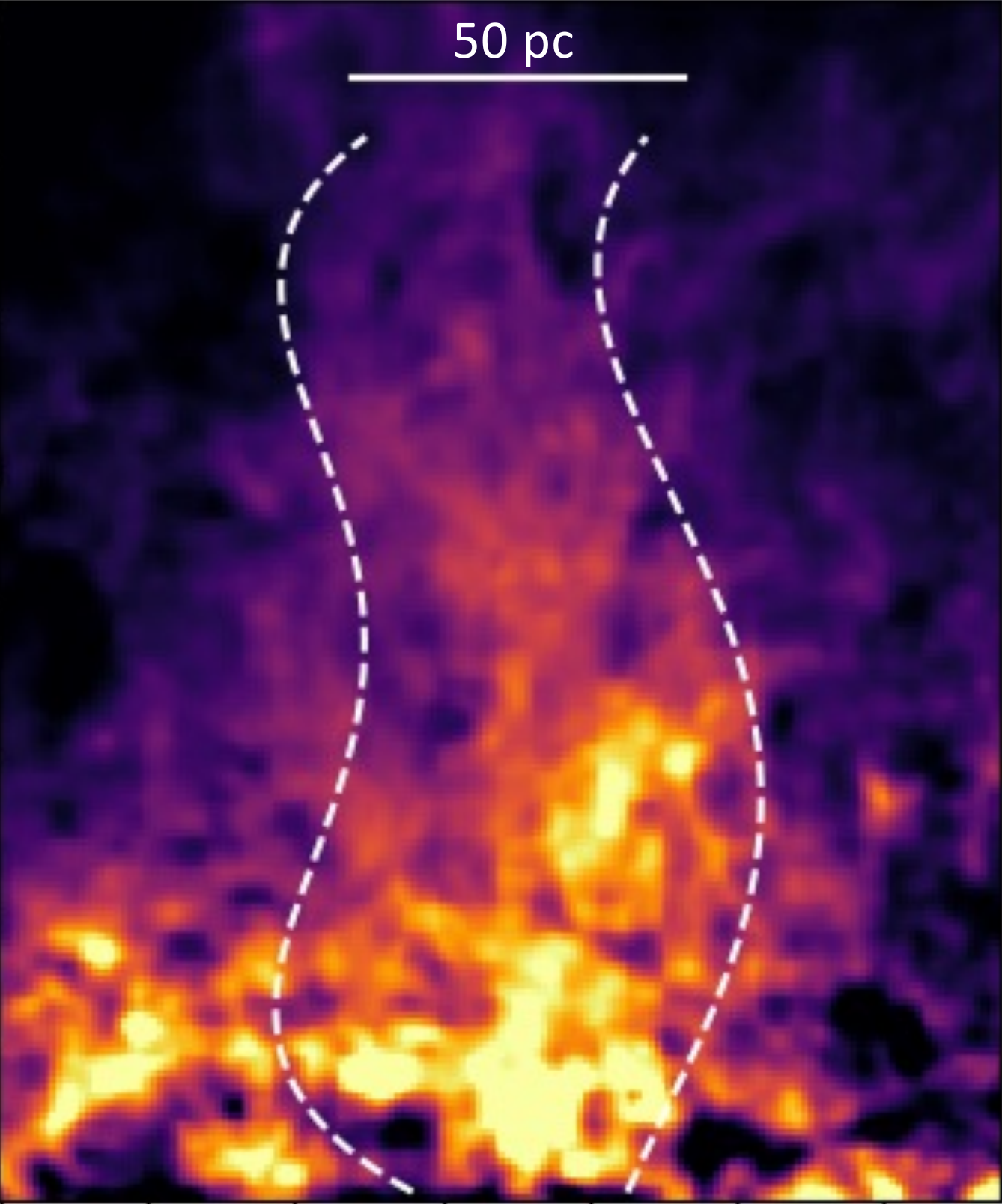}
    \caption{We show Plume-1, as an example of the plume shape and size determination. The local background emission has been subtracted. The white dashed lines represent the polynomial fit to the point-to-point width measurement, which truncates at the bottom of the image. The solid line at the top represents 50~pc.  }
    \label{fig:plume1}
\end{figure}

\begin{table}
 \begin{center}
 {\footnotesize\begin{tabular}{lcccc}\hline
 Plume & Side & Major Axis & Width & Width\\
         &      &    Position  &       & BKG Subtract \\
         &      &    [pc]  &    [pc]   & [pc] \\         
 \hline 
 1 & North &  -190 &  $72\pm10$ & $60\pm10$ \\
 2 & North & -43  &  $75\pm21$ & $56\pm20$ \\
 3 & North &  135 &  $84\pm25$ & $73\pm26$ \\
 4 & North & 236  &  $67\pm22$ & $49\pm32$ \\
 5 & South & -34  &  $50\pm10$ & $42\pm10$ \\ 
 6 & South & 92  &  $70\pm14$ & $57\pm19$\\ 
 7 & South &  219 &  $70\pm9$ & $54\pm17$\\  
 8 & South & 258  &  $45\pm9$ & $34\pm10$\\    
 \hline 

 \end{tabular}}
 \end{center}
 \caption{The properties of individual plumes identified. We give the position and width for each plume. We provide widths both without and with background subtraction (BKG Subtract). }
 \label{tab:chim} 
\end{table}

 \begin{figure}
     \centering
     \includegraphics[width=0.48\textwidth]{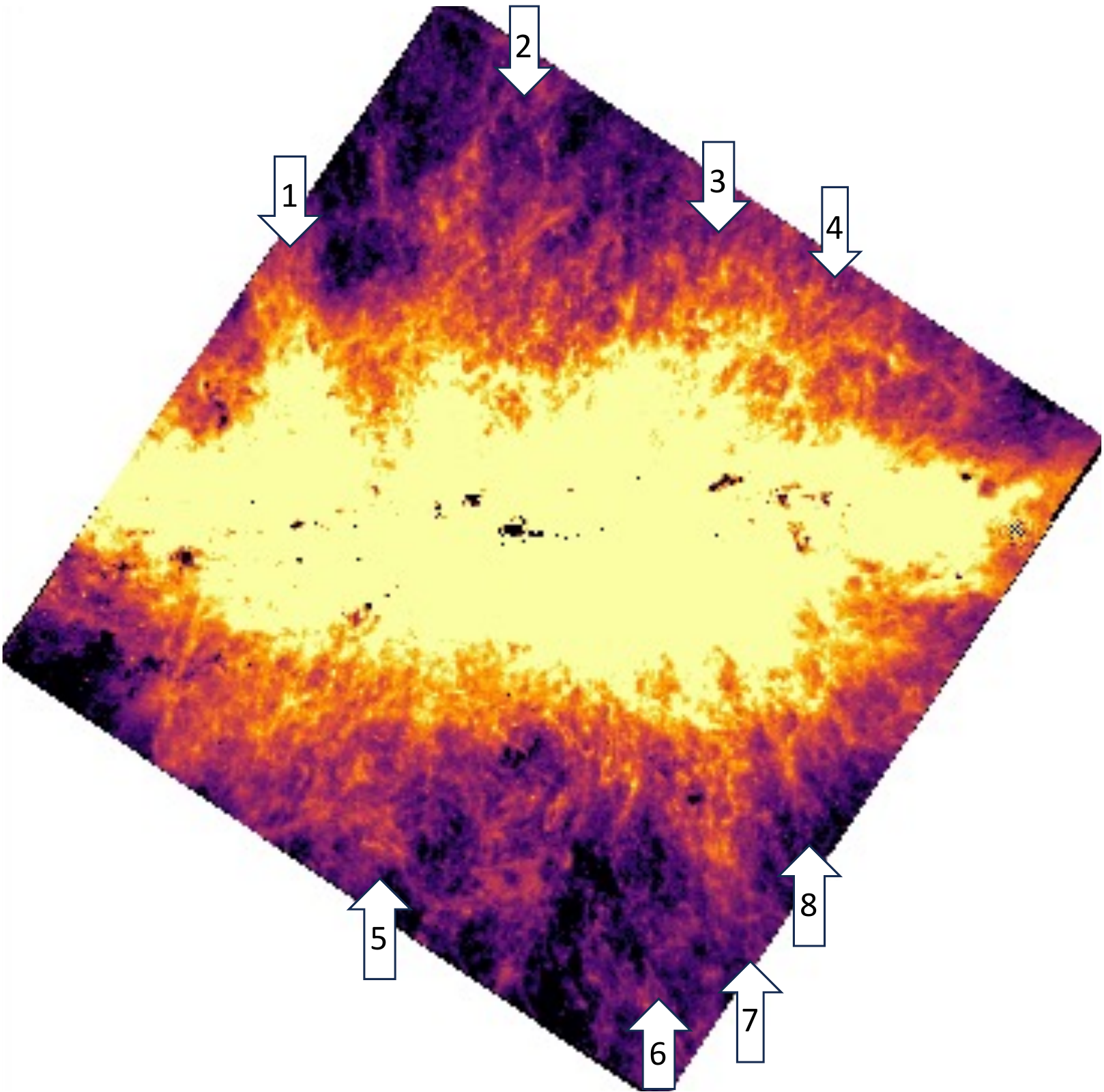}
     \caption{Here we show 3.3~$\mu$m PAH image again, but with the image scale set to square root and the contrast set to show the faintest substructure in the plumes. The locations of the plumes are denoted by arrows. The majority of the plumes extend to the edge of the field-of-view indicating that their length is greater than $\sim$300~pc.}
     \label{fig:stretch}
 \end{figure}

We now seek to measure the location, size, and shape of the plumes that we identified in Fig.~\ref{fig:pretty}.  We require a  minimum extent for plumes beyond the vertical half-light distance of the starlight in the F250M filter, $\sim$50~pc. This is intended to select significant plumes, though we acknowledge it is somewhat arbitrarily chosen. The plume does not need to be strictly perpendicular to the disk.

To measure plume properties we select a working window around each candidate. We then search for a significant peak contained within the central 3/4 of the window. This is done for every 0.1~arcsec region moving upward from the disk. If multiple upward steps fail to find a peak, then we identify this as a lack of coherence and the plume measurement is ended. Also, we interpret significant horizontal jumps in the location of the peak (greater than 1/3 the width) as a lack of coherent structure.  Once the peak location is identified we define a width. For each row we determine the horizontal distances from the peak that contain 2/3 of the flux in the window. We do not require that the sides of the plume are symmetric. We aim for the local working window, centered on the plume, to be as wide as possible. However, if the working window contains flux from a neighboring plume, this will impact the width measurement. Future work might investigate simultaneous decomposition of structures. This is however beyond the scope of this work. For each plume we iterate the boundary of the working window by $\sim$3 FWHM of the PSF of the image (20 pixels), and make the final selection on a region that has a stable solution for width and peak locations. After sides are determined for the whole plume we fit a 5$^{th}$ order polynomial to each side of the plume. This is done to smooth the point-to-point scatter in the width measurement. We then calculate the median width of each plume. The uncertainties are simply the root-mean-square deviation across each plume.

The choice to subtract a background of diffuse emission from the plume does impact the measured width. We therefore provide widths calculated with and without background subtraction. To calculate the background we measure the average flux of the first and last $1\arcsec$ in each row. On average, plumes are $\sim$14~pc wider without background subtraction. 

In Fig.~\ref{fig:plume1} we show Plume-1 (the left-most plume on the north side of the galaxy) as an example width calculation. In this case the median width is 60~pc (with background subtracted) and 72~pc with no background removal. This plume extends at least 200~pc, when the image field-of-view ends. 

For completeness, we also investigated peaks in Figs.~\ref{fig:northslice} and ~\ref{fig:southslice} that did not, ultimately result in a coherent plume. For example, in Fig.~\ref{fig:northslice} there is a peak near to  $+50$~pc. There is no coherent, continuous structure.  On the southern side of the galaxy there a number of peaks at major axis offset positions near $+$90~pc and $+$270~pc. When placing a window of similar size as others on these regions, however, it does not result in an independent plume. 

The average value of the median plume widths is 67~pc wide without background subtraction and 53~pc with background subtraction (see Table~\ref{tab:chim}). We note that this size scale is very well resolved, and not connected to any input parameters of measuring the plume widths. Plumes on the northern side of M\,82 are slightly wider than on the southern, with average widths of 59~pc and 47~pc for north and south (background removed), respectively. As discussed already the major axis offset positions are different, and we find that the southern plumes are less perpendicular to the galaxy major axis. The large scale asymmetry in the M\,82 outflow extends down to substructure at the  scale of tens of parsec. The plumes on the south side appear to be superimposed on top of a smoother distribution of gas. This is located in a similar region as the CO bubble described in \cite{Chisholm2016}. The slightly narrower size of the southern plumes could be due to not being able to separate the lower surface brightness flux from the underlying, more diffuse, gas.

Overall we find that 70\% of the flux beyond the central region ($\pm$50~pc) is associated with plumes. If the PAHs observed at larger vertical distance from the midplane of the galaxy \citep{Engelbracht2006} are transported from the disk, then these plumes we observe near the galaxy are the dominant lanes of PAH transport and possibly cold gas in general. The rest of the gas, away from the midplane, appears in a diffuse form. 

We find in our analysis that the flux from several of the plumes extends beyond the field-of-view of our NIRCam observations. In Fig.~\ref{fig:stretch} we replot the PAH emission with the flux scale displayed logarithmically, and the display levels are set to reveal fainter emission.  Plumes 1, 2, 4, 5, 7, and 8 extend to the edge of the field-of-view. Plume~2 appears  composed of multiple smaller scale filaments, which may be a bubble structure, near the edge of the field. Nonetheless a continuous structure can be traced to the edge of the field.  This implies that the length of plumes can reach at least to $\sim$300-400~pc, and likely extend into the larger wind. 

The location of PAH plumes and their peaks correlates with the location of peaks in other phases. We see that in Figs.~\ref{fig:northslice}~\&~\ref{fig:southslice}. There is a strong correlation of PAH peaks being co-located with Pa~$\alpha$ peaks on both the north and south side. In the horizontal flux slice of the chimneys (as indicated by grey bars) all are at or near peaks in the normalized Pa~$\alpha$. This similarity in Pa~$\alpha$ and 3.3~$\mu$m PAH was commented on in \cite{Bolatto2024}, and now quantified here. We will investigate the relationship between PAH and Pa~$\alpha$ further in a subsequent section. It is difficult to interpret CO in this small image only, as it is 17~times more coarse resolution. Nevertheless in the north outflow there is a connection between peaks in CO and the location of 3 out of 4 chimneys. The large values of CO at the edge of the horizontal surface brightness profiles are well known to be streamers associated with accretion \citep{Walter2002}. In the south side there is less correspondence with individual peaks,  although there is a correlation with the overall rise in CO flux and PAH flux from 0-200~pc in Fig.~\ref{fig:southslice}.


\section{Parsec Scale Clouds in M\,82 Winds}

\begin{figure*}[t]
\begin{center}
\includegraphics[width=\textwidth]{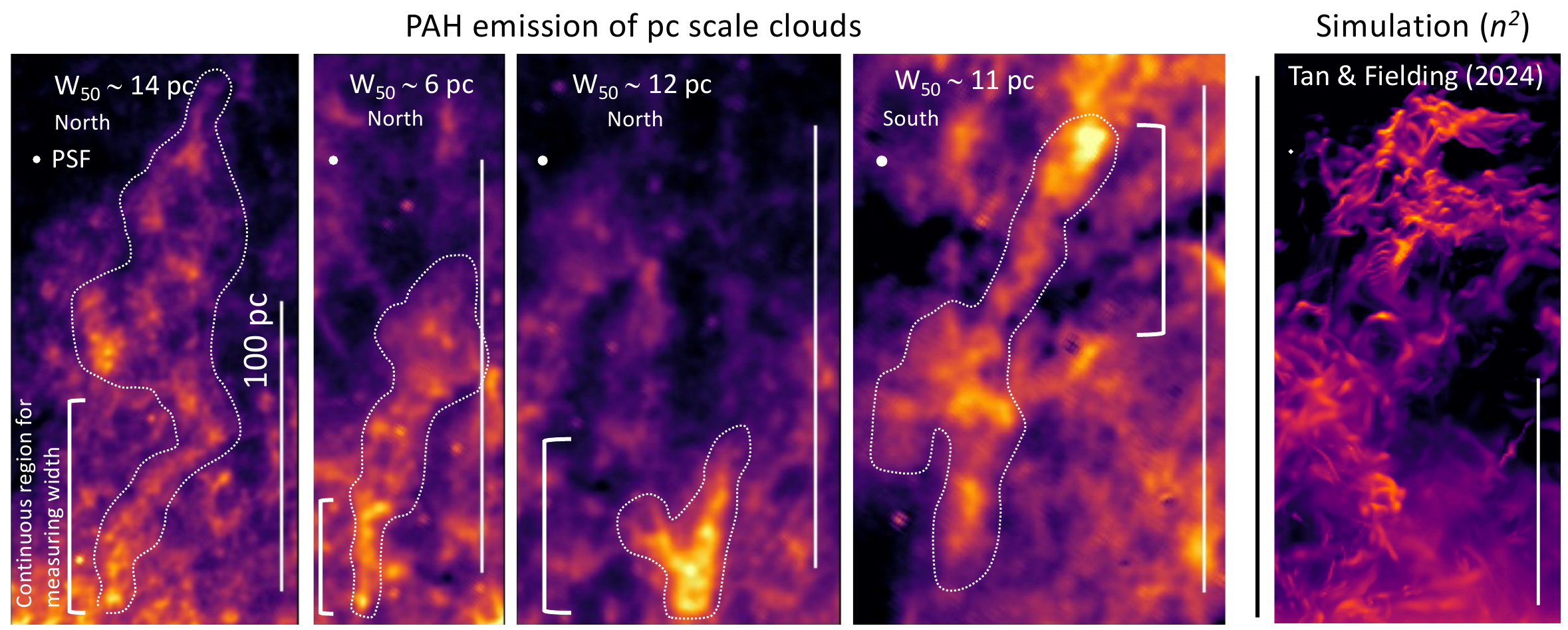}
\end{center}
\caption{ Cutouts of small-scale clouds in M\,82 winds. The white bar represents $\sim$100~pc, and the white dot in the upper-left indicates the PSF. The left three panels are taken from the northern side of the outflow, and the second-from-the-right panel from the southern side. The widths containing 50\% of the light are listed in the top. For each cloud we draw a contour that roughly encloses the cloud shape. The vertical bracket on the right side of each panel represents the distance over which the width is estimated. The right-most panel shows a simulation from \citet{Tan2023} of a cloud in an outflow. The figure shows density squared. The morphology of the emission is similar to what we observed in M\,82. } \label{fig:clouds} 
\end{figure*}

\subsection{Identification of Clouds}
The substructure of the 3.3~$\mu$m PAH emission in the M\,82 wind is clearly hierarchical. The plumes we describe in the previous section are comprised of a smaller scale substructure.  \cite{Bolatto2024arXiv} show evidence of this via unsharp masked imaging. We will refer to these smaller features as clouds. The distinction between the clouds and plumes (discussed in the last section) is that multiple clouds may exist within a single plume.

In Fig.~\ref{fig:clouds} we show examples of small-scale clouds of 3.3~$\mu$m PAH taken from the plumes in M\,82. All of these are located in plumes discussed in the previous section. A common feature of these clouds is a higher surface brightness knot of emission that is followed by a lower surface brightness emission extending away from the galaxy. The clouds are stretched in the vertical axis. A large number of such structures are visible in the M\,82 wind, many of which are overlapping in a network, or web-like.  At the low surface brightness end of the cloud the distinction between diffuse gas of the plume and the cloud is likely arbitrary. The bright knot typically has 2-4 times the surface brightness of the tail. These outflow clouds can have extents longer than $\sim$30-150~pc. 

Our imaging data cannot distinguish if the overlapping structures are truly connected, or if they are unconnected structures superimposed on top of each other. Both scenarios could be true. The latter scenario should be common if the total outflow is arranged in a biconical structure. 

A number of simulations in recent years predict that small-scale clouds of cold gas can exist within outflows in a range of environments \citep{Fielding2018,gronke2018,Sparre2020,Tan2023,Abruzzo2022, Zhang2024}. Of particular interest to this work is \cite{Farber2022}, \cite{Chen2023} and recently \cite{Richie2024arXiv}. These include rough self-shielding approximations that allow for the formation of colder phases (1000~K and 100~K, respectively), which makes them appropriate for comparison to our  PAH observations. These simulations adopt a ``cloud-crushing'' set up in which an over-dense sphere of cold gas is hit by a hot (${\sim} 10^6$~K) wind. The general finding is that if cloud size and density are sufficiently large, then the cloud survives the shock and is accelerated by the wind. In the wind, clouds typically stretch into a comet-like structure, with a bright base pointing toward the source of the wind and an extended filamentary tail. The width of the cloud, at least near the base, is similar to the initial cloud diameter. Qualitatively speaking, this structure is what we observe in Fig.~\ref{fig:clouds}.

In Fig.~\ref{fig:clouds} we directly compare these small clouds observed in the M\,82 outflow to a collection of clouds in the multiphase wind launched in the ISM patch simulation presented in \cite{Tan2023}. They study cold clouds that emerge from a simulated superbubble in a box with a size of 512$\times$512$\times$2048~pc$^3$. This simulation sits between the idealised ``cloud-crushing'' simulations and large-scale full galaxy simulations. They have a comparable spatial resolution to our observations, $\sim$1~pc. The initial conditions of the launching mechanism and disk include clustered supernovae, as would likely happen in the massive star clusters in M\,82, coming from a section of a gas disk that has $\Sigma_{\rm g}\sim175$~M$_{\odot}$~pc$^{-2}$ . This is a similar, though perhaps slightly lower, surface density to that observed for M\,82 \citep{Leroy2015,Krieger2021}. 

Qualitatively speaking, the size and morphology of the 10~pc scale outflow clouds of PAH we observe in M\,82 are similar to the clouds simulated in \cite{Tan2023}. We note that \cite{Tan2023} argue that their simulations of clouds validates the qualitative results of the cloud crushing simulations, as well as the more quantitative predictions from detail mixing layer studies \citep{Fielding2020, Tan2021}. 

\begin{figure}
    \centering
    \includegraphics[width=0.5\textwidth]{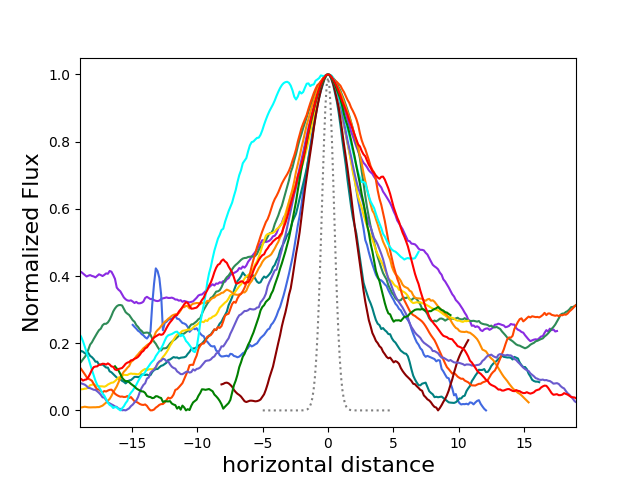}
    \caption{The width of 12 clouds of PAH emission is shown. The colors are arbitrary, so that one can follow individual clouds. The dotted line represents a Gaussian with FWHM$\sim$1.9~pc, which is the nominal spatial resolution. We measure 50\% cloud widths of 5-18~pc. }
    \label{fig:cloudprofs}
\end{figure}

The spatial resolution of our NIRCam 3.3~$\mu$m images allows us to measure sizes larger than $\sim$2~pc in M\,82. The white dot in each panel of images in Fig.~\ref{fig:clouds} shows the PSF size. The substructure is marginally resolved. We therefore aim to directly measure the width, which can be related to the physically important $R_{\rm cloud}$. 
To our knowledge these are the first direct estimates filament widths for clouds embedded in a wind outside of the Milky Way. Other existing estimates of cloud sizes \citep[e.g][]{Xu2023} rely on modelling of line ratios. 

In order to measure the cloud sizes we first identify 12 clouds by visual inspection of the full resolution 3.3~$\mu$m PAH image. Due to the morphological differences in the northern and southern outflow we identify 6 clouds on the northern outflow and 6 on the southern. We avoid clouds that show significant overlap with other clouds. We also restrict our analysis to clouds that begin at roughly $\pm$200~pc from the midplane of M\,82. 

We caution the reader that these selection criteria almost certainly introduce biases into our sample of clouds. For example, if clouds grow via collisions with other clouds, that may be missed by our selection as they may have more complex structure. Alternatively, there may be a bias in the types of clouds that exhibit cometary structure. The numbers we present should be taken as a first estimate of what is possible. A full analysis of the entire outflow is planned by our team.

We measure cloud widths in a procedure not unlike that measuring the chimney widths. We first define a region that includes the cloud (those are shown in Fig.~\ref{fig:clouds}), and a starting position at the base of the cloud. The software then finds the peak flux in each row. If the position of the peak jumps by more than 3~PSF FWHM from the previous position the software stops. A normalized surface brightness profile of each row is then determined, with the peak position at $x=0$~pc. We then take the median of all the resulting rows. Finally the width of each cloud is taken to be the width that contains 50\% of the PAH emission in the normalized median cloud profile. We refer to this width as $W_{50}$. We show all 12 normalized profiles in Fig.~\ref{fig:cloudprofs}. The largest width we measure is $\sim$18~pc and the smallest $\sim$5~pc. There is not a noticeable difference between north and south, though our sample is probably too small to identify any small scale trend. The average width of all twelve clouds is $\sim$9.8~pc. 

We can compare these clouds to the cold clouds observed in the fountain of the Milky Way \citep[e.g.][]{diTeodoro2020,Noon2023}. \cite{Noon2023} use new, high resolution HI imaging from MeerKAT of 3 clouds in the Milky Way. They find widths that are similar to ours, $\sim$5-10~pc. \cite{Xu2023M82} uses optical emission line ratios to derive cloud sizes in M\,82. We note that this calculation requires a significant number of assumptions. Moreover, \cite{Xu2023M82} make this calculation for optically-emitting ionized gas, a different phase of gas that may have different widths from the cold phase studied here. They derive sizes that are $R_{\rm cl}\sim$1~pc. This is smaller than our PAH widths. We do not know, observationally, how the sizes of clouds would change with different phases, although inspection of simulations \citep[e.g.][]{Farber2022,Chen2023,Richie2024arXiv} would suggest that the warmer ionized gas would surround the cold gas. Moreover, we state again that our cloud selection may be biased, and may favor larger clouds. Future work, by our team, on the full M\,82 outflow imaged at 3.3~$\mu$m will allow for a more complete study of cloud widths. Chastenet et al. {\em in prep} are studying similar clouds in the galactic fountain of NGC~891. JWST is clearly an excellent tool for studying the small-scale structure of outflow gas. More observations of a range of outflows is now needed to understand the full range of these building blocks of outflows.

\subsection{Timescales for Survival of Clouds }

Theory suggests that the so-called ``cloud crushing time'' is a critical parameter to determine the survivability of clouds like those we observe in Fig.~\ref{fig:clouds}. This can be estimated as $t_{\rm cc} \sim \chi^{1/2} R_{\rm cloud}/v_{\rm rel}$, where $\chi$ is the original density contrast of the cloud to the wind fluid, $R_{\rm cloud}$ is the radius of the cloud, and $v_{\rm rel}$ is the velocity of the cloud relative to the wind fluid. We do not know the original properties of the clouds, but many simulations suggest it is appropriate in wind environments to adopt $\chi\sim100-1000$ \citep{gronke2018,Sparre2020,Abruzzo2022}. For M\,82 the asymptotic velocity of the hot wind is roughly $v_{\rm hot}\sim1000$~km~s$^{-1}$ \citep{Strickland2009}. We do not know the velocity of the PAH-emitting gas, but we can assume that it moves at a similar velocity to either the ionized gas or the molecular gas. In M\,82 ionized gas is known to have $v_{\rm H\alpha}\sim 500$~km~s$^{-1}$ \cite[e.g.][]{Shopbell1998,Xu2023M82}. Molecular gas is somewhat slower, observed to have velocities $v_{\rm mol}\sim200-300$~km~s$^{-1}$ \citep{Leroy2015,Krieger2021}. Given the strong correlation between PAH and Pa~$\alpha$, we take $v_{\rm rel}\sim 500$~km~s$^{-1}$. If the PAH-emitting gas has velocities similar to the gas observed in CO this will increase $v_{\rm rel}$ and thus decrease $t_{\rm cc}$ by roughly a factor of two. Alternatively, if we take the initial velocity of the cloud as the appropriate characteristic this would be significantly lower, $\sim$10~km~s$^{-1}$, which would imply a higher $v_{\rm rel}\sim 1000$~km~s$^{-1}$ and thus lower $t_{\rm cc}$.  For the size of the cloud, we simply take $R_{\rm cloud,PAH}\sim 2\times (W_{50}/2)\sim10$~pc. This yields $t_{\rm cc}$ of order $\sim$0.3-3~Myr. 

We can likewise estimate the time needed for the cloud to travel to its current position. The cometary heads of clouds in Fig.~\ref{fig:clouds}~\& \ref{fig:cloudprofs} are roughly 200~pc from the disk midplane. For this we assume the PAH has roughly similar velocity as the $H\alpha$. This translates to a cloud lifetime of $\sim$0.5-1~Myr. Note if the cloud launches from a higher position than the midplane then lifetime is shorter. The cloud-crushing time is therefore equal to or larger than the lifetime of the cloud, if it originated at the midplane. These timescales are therefore consistent with clouds originating as entrained gas and surviving to this point. We note that our by-eye selection biases our measurement to clouds that, by selection, still exist. An automated analysis of all substructure may find fainter, smaller structures with morphology suggestive of later stages of destruction. 

The basic criteria that cooling time of mixed gas is smaller than that of the cloud-crushing time is easily met for M\,82. Assuming $n_e\sim100$~cm$^{-3}$ \citep[e.g.][]{Xu2023M82} for 10$^4$~K gas the typical pressure is $P/k_b\sim10^6$~cm$^{-3}$~K, which corresponds to $t_{\rm cool,mix}\sim10^{-4}$~Myr \citep{Abruzzo2022}. Thus these cloud sizes are more than sufficient to survive the disk breakout. 
Likewise, \cite{Bolatto2024} discuss the time-scale for PAHs to be destroyed in M~82 wind. Calculations for collisional destruction of PAHs suggest that those time scales are long, in excess of 10~Myr \citep{Micelotta2010}. Photodestruction time scales are very uncertain, particularly for PAHs embedded in ionized gas, but they may be considerably shorter. PAHs in the ionized gas, however, may be constantly replenished as neutral gas is ionized, for as long as a neutral or molecular gas reservoir exists. 

Cloud sizes are very important to establish $t_{\rm cc}$, and we could have underestimated those in our calculations above. A plausible scenario is that the cold gas we trace with PAH is surrounded by warmer ionized gas. The majority of simulations and theory surrounding cloud crushing focuses on gas that is above $10^4$~K, which would be more likely ionized. Those simulations that do consider low temperature gas \citep{Farber2022,Chen2023} show temperature gradients such that the coldest gas is concentrated in the middle of the cloud. This would imply a larger $R_{\rm cloud}$ and thus a larger $t_{\rm cc}$. Also, the simulations of both \cite{Chen2023} and \cite{Farber2022} consider pressures that are multiple orders of magnitude smaller than what is observed in M\,82. This generates clouds that are larger, following \cite{Abruzzo2022}. In principle, we expect that if simulations of cold gas are run that match the conditions of M\,82 they may indeed generate cloud sizes more similar to ours. Alternatively, if we have underestimated $v_{\ rel}$, either from underestimating the velocity of the hot wind or overestimating the velocity of the cold gas, then this would reduce $t_{cc}$, and suggest that these clouds should be more rare at larger distance. 

\begin{figure*}
\begin{center}
\includegraphics[width=\textwidth]{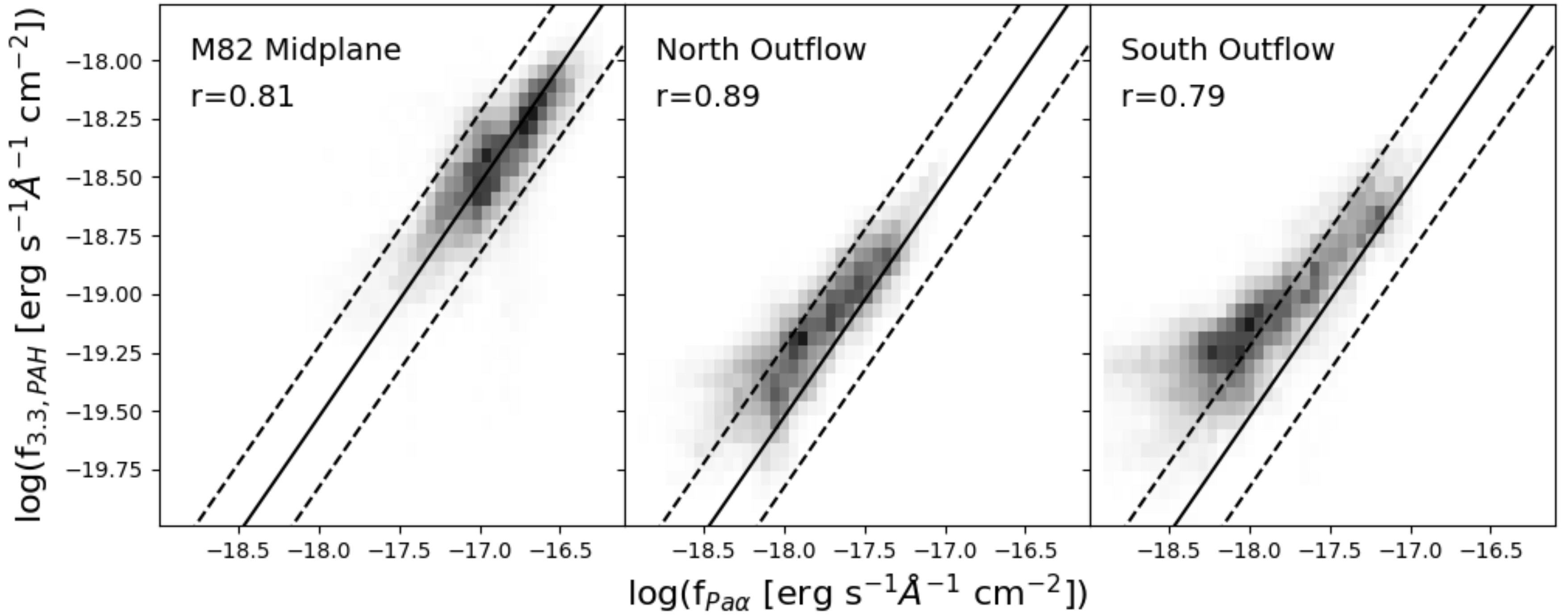}
\end{center}
\caption{ The relationship between PAH flux and Paschen~$\alpha$ is shown for the center of M\,82 (left), the northern outflow (middle) and the southern outflow (right). In each plot we show the relationship derived for the center of M\,82 as a black solid line. In all panels there is a strong correlation between the gas tracers. There is an elevation of PAH flux compared to Paschen~$\alpha$ in the outflow regions.    \label{fig:pahvspaa} }
\end{figure*}

\begin{figure}
    \centering
    \includegraphics[width=0.49\textwidth]{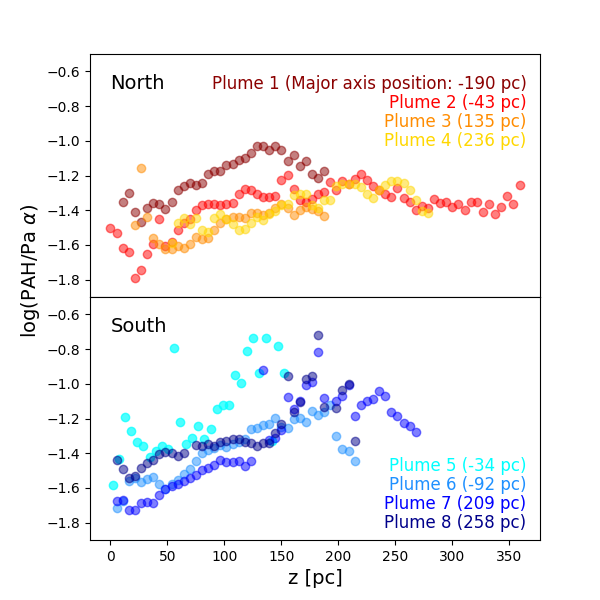}
    \caption{The PAH/Pa~$\alpha$ flux ratio for each of the plumes identified in section 3 is plotted the distance to the midplane of the galaxy. North and south outflows are plotted in top and bottom panels respectively. We see an increasing ratio with distance from the galaxy. }
    \label{fig:pah2pasch_profile}
\end{figure}

Overall, we take these results as encouraging for the picture of cloud entrainment and survival. First, the morphological similarity of our clouds to those produced in cloud-crushing simulations suggests that the overall concept may be valid. Moreover, we have demonstrated that the cloud properties observed in the M\,82 wind are consistent with what theory predicts is required for a cloud to survive breakout from the disk without being shredded by the hot wind. 



\section{Correlation of PAH and Pa~$\alpha$ in both Midplane and Outflow}

In section 3 we discuss the connection between the location of the plumes in PAH and the colocation of peaks in the Pa~$\alpha$ image. In Fig.~\ref{fig:pahvspaa} we investigate this further showing the pixel-to-pixel correlation of 3.3~$\mu$m PAH emission with that of Pa~$\alpha$. The 3.3~$\mu$m PAH feature is generally understood to arise from small, neutral PAHs \citep{Draine2021}, and thus does not {\em a priori} require strong correlations with ionised gas. Indeed, recent observations of the Orion Bar, with much finer (milliparsec) resolution, show that 3.3~$\mu$m emission decays where Pa~$\alpha$ is brightest \citep{Peeters2023}. 

For our analysis the PAH image is convolved and resampled to match the NIC3 F187N observations. The FWHM of NIC3 is $\sim$0.35~arcsec ($\sim$6~pc). This is sufficient to identify the plumes. At the small scale cloud substructure, roughly half the cloud would be covered by a single resolution element. It is, therefore, more difficult with this spatial resolution to separate a cloud from the background emission. 
\begin{figure*}
    \includegraphics[width=\textwidth]{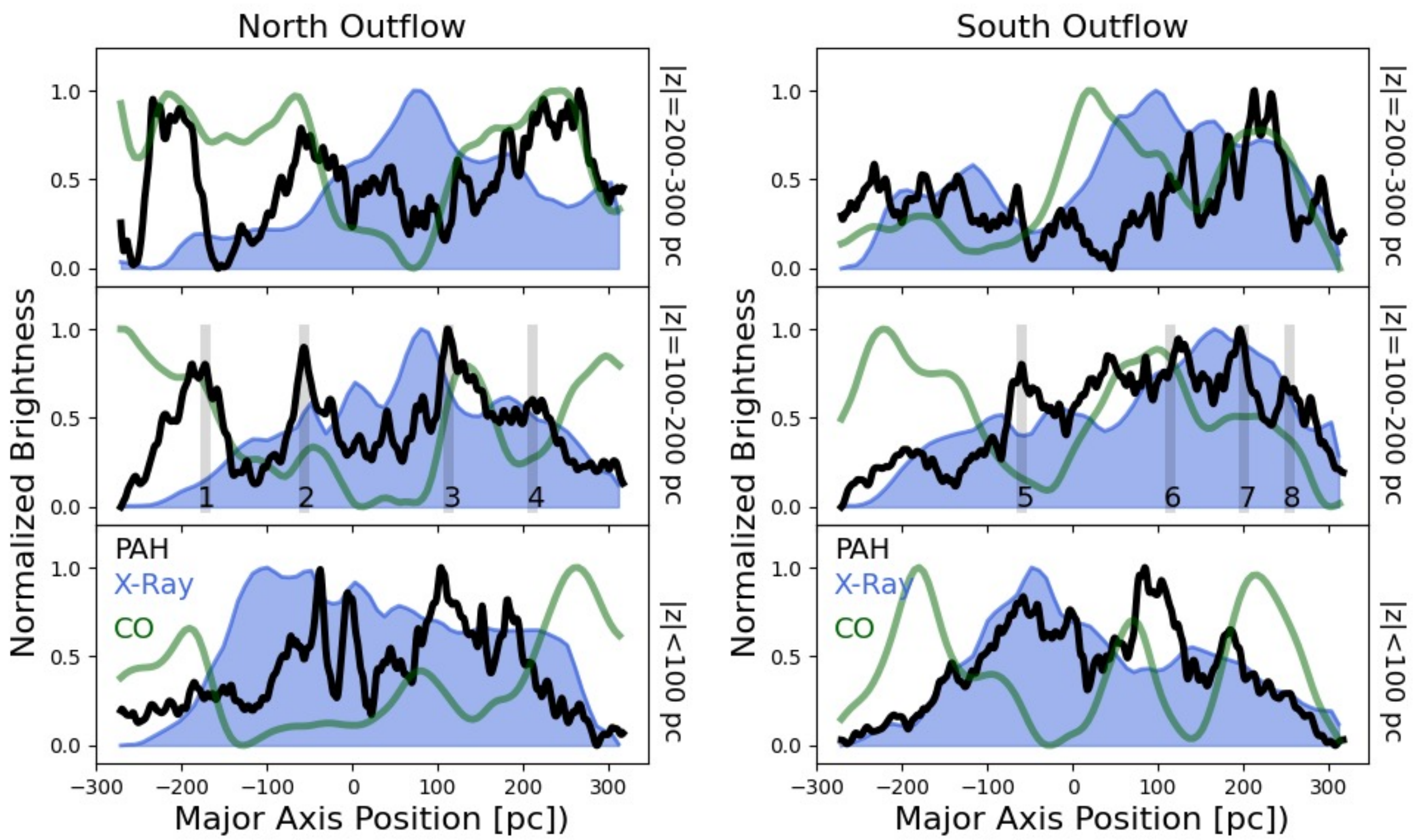}
    \caption{Similar to Fig.~\ref{fig:northslice} we show the normalized horizontal profile of the PAH flux (black) and the CO(1--0) flux (green). In this figure we add the X-ray flux (blue) from \citet{Lopez2020}. The middle panel is a similar region as in Fig.~\ref{fig:northslice}~\&~\ref{fig:southslice} on the left and right respectively. The bars in the middle panel show the location of the PAH plumes. We also show the galaxy center and a region at higher elevation than the middle panel. These profiles establish that the well-known picture in which cold gas surrounds a hot wind establishes itself within a few hundred parsec, and is important for setting the morphology of the PAH plumes.}
    \label{fig:xray}
\end{figure*}

In Fig.~\ref{fig:pahvspaa}, we show that there is a very strong correlation between PAH and Pa~$\alpha$ emission in both the midplane and outflow of M\,82. We find Pearson's correlation coefficients of $r=0.8-0.9$ for each component of this galaxy. In the galaxy midplane there is $\sim$0.2~dex scatter around a linear relationship.   Similarly, \cite{Belfiore2023} studies 3.3~$\mu$m emission toward HII regions in PHANGS galaxies, with spatial resolution of order $\sim$15-50~pc \citep[see also review by][]{Schinnerer2024arXiv}. The 3.3~$\mu$m feature does not experience as strong a PAH decrement in bright H$\alpha$-emitting HII regions as other PAH features show \citep[e.g.][]{Gordon2008,Egorov2023}. Our strong correlation in the midplane of M~82 is, therefore, not unlike the behaviour of HII regions in the PHANGS survey.  

In the middle and right panels of Fig.~\ref{fig:pahvspaa}, we find that the relationship between 3.3~$\mu$m PAH and Pa~$\alpha$ remains strong ($r\sim0.8-0.9$) in the outflow. There is, however, an increase in f$_{3.3}$/f$_{\rm Pa\alpha}$ in both the northern and southern outflows relative to the relation observed in the midplane. This ratio also increases for lower PAH and Pa~$\alpha$ fluxes. We find that in the midplane of M\,82 log(f$_{3.3}$/f$_{\rm Pa\alpha})\sim-1.6$. This increases in the wind to $\sim$-1.3 (north side) and $\sim$-1.15 (southside). The relationship between 3.3~$\mu$m PAH and Pa~$\alpha$ in the north outflow remains closer to linear, while in the south there is shallower a slope of f$_{3.3}$ with f$_{\rm Pa\alpha}$. 

There is a dependence in both the observed PAH and Pa~$\alpha$ fluxes with distance from the midplane, and therefore decreasing flux is essentially equivalent to increasing distance from the midplane within our field-of-view. We show this in Fig.~\ref{fig:pah2pasch_profile}. Here we plot the f$_{3.3}$/f$_{\rm Pa\alpha}$ for each plume identified in section 3 of this paper. In all plumes the  f$_{3.3}$/f$_{\rm Pa\alpha}$ ratio increases from the midplane to a position of $|z|\sim 100-150$~pc from the disk. Only a few plumes are measurable beyond this. Plumes 1, 2, 4 and 7 seem to no longer show an increase in f$_{3.3}$/f$_{\rm Pa\alpha}$ beyond this position. 

We did not apply an extinction correction to Paschen~$\alpha$, as it is difficult to make pixel-to-pixel estimates in the midplane of the highly extincted M~82, with studies finding variation in the disk of $A_V \sim 2-12$~mag \citep[e.g.][]{Forster2001}.  The extinction has been observed to decrease as a function of distance from the galaxy \citep{Westmoquette2013}. The impact on the vertical gradient, however, would then be to steepen this gradient. 

Overall, we can place the observed relationship of 3.3~$\mu$m~PAH and Pa~$\alpha$ in the context of recent observations of spiral galaxies. In local spirals from the PHANGS survey observations show tight relationships between extinction corrected H$\alpha$ and PAH emission lines \citep{Belfiore2023,Leroy2023,Schinnerer2024arXiv}. This is very similar to our relationship in Fig.~\ref{fig:pahvspaa}. 

We find that in the central starburst of M~82 that $L_{PAH}(F335M)\approx 17 L(Pa \alpha)$. It is difficult to interpret any comparison this to similar results of correlatiosn of PAH features with Balmer lines \citep[e.g.][]{Leroy2023,Belfiore2023,Schinnerer2024arXiv}. The central starburst of M~82 is very different than HII regions, which the PHANGS measurements are based on.  Our ratio of $L_{PAH}(F335M)/L(Pa\alpha)$ is lower by a factor of $\sim$5-6 compared to the PHANGS. The low value of our number may indicate a larger amount of diffuse ionised gas along the line-of-sight. The orientation of M~82 likewise complicates this comparison, as PHANGS galaxies are face-on and M~82 is edge-on. Further investigation of starburst galaxies will be informative. Independent of any decrement in PAH emission, when Fig.~\ref{fig:pahvspaa} clearly demonstrates is a strong correlation of PAH and ionised gas flux in a broad range of galaxy types.  
 
Based on observations PHANGS galaxies, \cite{Leroy2023} argues that the emission of mid-IR features, like 3.3~$\mu$m PAH, is driven by a mixture of heating (traced by ionised gas) and column density of cold gas. Our results in Figs.~\ref{fig:pahvspaa}~\&~\ref{fig:pah2pasch_profile} can be understood in this framework. The tight correlation of Pa~$\alpha$ and PAH in the midplane of M~82 suggests that heating is dominating the emission. As we move to larger distances from the galaxy the heating sources decrease. This is observed as a decrease in the median Pa~$\alpha$ surface brightness by roughly an order of magnitude in the outflow compared to the midplane. This is similar to PHANGS observations \citep{Leroy2023}, where gas with fainter H$\alpha$ emission has higher ratios of PAH/H$\alpha$, in that case using the 7.7~$\mu$m and 11~$\mu$m emission. As the ionised gas flux decreases the relative importance of cold gas column density increases.  Future work using the larger field-of-view observations of M~82 and more PAH emission lines will be informative to compare to this model.

\section{PAH Plumes Are Suppressed in Bright X-Ray Regions}

In Fig.~\ref{fig:xray} we compare the horizontal flux profiles of 3.3~$\mu$m PAH and CO(1--0) to the X-ray emission \citep[taken from ][]{Lopez2020}. A commonly adopted picture of the multiphase structure of outflows is one in which the hot X-ray gas establishes a central lobe that is surrounded by colder gas \citep[e.g.][]{Leroy2015}. Our JWST NIRCam observations allow us to resolve the relation between PAH-emitting cool gas and X-ray emitting hot gas on small scales. 

The bottom panel of Fig.~\ref{fig:xray} shows that in the midplane of M\,82 X-Ray and PAH emission overlap. There is a strong X-ray peak at major-axis position -50~pc, which corresponds to a local peak in PAH emission. This behavior changes in the outflow, although the pattern is somewhat different on the north and south sides of the galaxy. We will, therefore, describe these separately.  

{\bf North Outflow:} On the north side of M\,82 (Figure \ref{fig:xray} left) the separation between X-ray and PAH emission becomes apparent within 100~pc of the galaxy midplane. The brightest X-ray peak in this z-axis elevation is located at a major axis offset of $+80$~pc from the galaxy center, with emission falling off at larger offsets. Two PAH plumes surround this central peak, located at roughly -55~pc and $+$110~pc offsets. There is a trough in the PAH emission between these two plumes, the location of which corresponds to the X-ray peak. As we have shown in Fig.~\ref{fig:northslice} the peaks in CO(1--0) are spatially correlated with the PAH. Now we see that both CO(1--0) and PAH are anti-correlated to the X-ray peak. The rest of the PAH emission plumes are found at larger major axis offsets from the X-ray peak. 

At the farthest distance from the starburst ($z\sim200-300$~pc), both the PAH and CO emission completely avoids the X-ray peak. Plume-3 at $z\sim 100-200$ (located closest to the X-ray peak, at offset $+110$~pc) is no longer prominent, although there is still a local peak at the offset position of plume-3.
Prominent PAH emission plumes on the north side of the outflow are at offsets of $\sim150-200$~pc from the X-ray peak.  At this distance above the plane, the CO has its minimum at the same location as the X-ray peak. 

{\bf South Outflow:} The separation between X-ray and PAH emission does not develop as quickly in the south outflow as in the north. In the $z\sim100-200$~pc range on the south the peak in X-ray emission is located at a major axis offset of $+167$~pc from the galaxy center. This peak is broader than the peak on the northern side. The X-ray, CO and PAH emission peaks on the south side all overlap.
We noted above that the southern plumes are less perpendicular to the galaxy major axis, closer together and less prominent than those on the north. The coincidence with the hot wind may be related to this. 

At larger distances from the galaxy mid-plane ($z\sim200-300$~pc) the southern outflow begins behaving more like the north. The X-ray peak has moved to $\sim$100~pc and is narrower. The PAH emission no longer peaks at the same location as the X-ray. The most prominent PAH peak is at an offset of $\sim$+210~pc, at least 100~pc away from the X-ray peak. This PAH peak is colocated with a CO peak.  Because the southern PAH plumes diverge away from the galaxy center, Plumes 7 \& 8 are therefore, avoiding the X-ray peak at this height. There is another CO peak at 20~pc, that likewise avoids the X-rays, though is not colocated with a local PAH peak. 

In both the north and the south outflows by $z\sim200-300$~pc there seems to be an established separation between the X-ray and the PAH emission. On the north side this happens as early as $\sim$100~pc from the midplane. Previous work \citep[e.g.,][]{Leroy2015,Engelbracht2006,Strickland2009} shows this behavior continues to larger distances. We note that because M\,82 is highly inclined we do not know the relative position of the PAH plumes and X-ray emission along the line-of-sight. A plausible scenario on the north side of M\,82 is that plumes 2 \& 3 (the two plumes closest to the galaxy center) are located further from the galaxy center along the line-of-sight. In this case the cold gas would surround the hot wind. 

The observation that plumes avoid the bright X-ray regions of the outflow has implications for the small scale clouds observed in Fig.~\ref{fig:clouds}. Increasing the temperature of the wind means that the minimum cloud size that survives will likewise increase \citep{gronke2018}. If those clouds are located in plumes of cold gas it is plausible that the surrounding gas is cooler. The results of Fig.~\ref{fig:xray} suggest that indeed the PAH resides in locations of cold gas, and that these structures avoid regions of X-ray emission.  The strong correlations of PAH and Paschen~$\alpha$ at scales of $\sim$5~pc can be understood if the PAH emission is coming mostly from gas at the PDR, close to the photo ionized gas (which itself is only 3000-10000~K, still much cooler than the X-ray emitting gas). The combination of CO spatial resolution and PAH field-of-view make establishing scaling relations, similar to Fig.~\ref{fig:pahvspaa}, between CO and PAH difficult with the data in this paper. The hierarchical structure of small scale clouds, residing in plume super-structures may be necessary to transport cold gas from the disk to the outflow. The 3.3~$\mu$m PAH feature is the best high-spatial resolution tracer of cold gas in the diffuse environment of outflows. JWST observations of 3.3~$\mu$m PAH in more galaxies with galactic winds are clearly needed. 

\section{Summary and Conclusions}

The 3.3~$\mu$m PAH emission ejected from the starburst of M\,82 is not a smooth structure, it is rather organized in plumes, and those plumes are composed of smaller scale clouds (some with identifiable cometary shapes).
We find that the plumes have widths of order $\sim$50-75~pc. We identify 4 coherent plumes on each side of the galaxy. At least one plume on the north side merges with adjacent emission, and there are other structures on the south side with similar widths that are shorter. The identified plumes comprise 70\% of the 3.3~$\mu$m flux. We find that most of the plumes extend to the edge of our field-of-view. We suggest that the plumes may be the dominant method by which cool gas is transported to large distances from the starburst by the wind. 

These plumes, therefore, should be reproduced by simulations of winds that are intended to describe outflows like M\,82. For example, recent simulations show that the impact of cosmic rays is to smooth out the clumpy substructure \citep[e.g.][]{Rathjen2021}. Our observations may indicate that while cosmic rays may play an important role in Milky Way-like environments \citep{Armillotta2022}, they are less important when the starburst is as strong as in M\,82. This can be test with observations of JWST on nearby edge-on systems with low star formation rates. 

We find that the locations of the PAH plumes are likewise the locations of brighter Pa~$\alpha$ peaks \ref{fig:northslice} \& \ref{fig:southslice}. Moreover, we show that in pixel-to-pixel comparison there is a tight correlation of 3.3~$\mu$m PAH flux with Pa~$\alpha$ flux. Our interpretation of this is that the plumes in PAH emission are also likely substructures in ionized gas, likely arising at the photoionized surfaces of neutral or molecular clouds. Comparison to CO emission is hampered by the lower spatial resolution and sensitivity of the available data, although there do seem to be local peaks in CO that are co-located with the PAH emission plumes. \cite{Bolatto2024arXiv} show that these plumes also are co-located with bright plumes in 6 GHz radio continuum, caused by free-free emission. The bright radio continuum regions were first shown by \cite{Wills1999}. In future work we plan to make a detailed comparison to the VLA data. 

We show in Fig.~\ref{fig:clouds} that the plumes are comprised of smaller clouds that often have an elongated structure with a bright base in the direction of the starburst. We note the similarities between this cometary shape and those results of cloud-crushing simulations \citep[e.g.][]{gronke2018,Sparre2020,Armillotta2017,Abruzzo2022,Farber2022}. We measure the widths of 12 clouds selected from the image, finding a typical width ${\sim}5-18$~pc, with an average of 9.8~pc. Taking this to indicate the cloud size we estimate the cloud crushing timescale $t_{\rm cc}\sim 15$~Myr. This changes by a factor of $\sim$2  depending up assumptions of initial cloud overdensity and wind speed. Our selected clouds are all taken from distances $z\sim$200~pc from the midplane. We find that $t_{cc}$ is very similar to the travel time of the cloud from the galaxy midplane, assuming a velocity of order 500~km~s$^{-1}$. The general prediction of simulations is that absent stabilisation mechanisms, such as cooling, cold clouds are destroyed on a timescale of a few $t_{\rm cc}$. We selected clouds that were fairly isolated and with sufficient signal-to-noise to measure the size, which possibly biases us to more prominent, possibly more intact, clouds. Nevertheless, these sizes are consistent with a picture in which the cool clouds entrained in the outflow survive for some time in the  wind. 

A straightforward interpretation of the plumes extending outward from the disk and the viability of cloud survival timescales is that the clouds we observe at 200-300~pc in M\,82 are entrained in the hot flow within the disk and  transported outward. We take these results as strongly encouraging further work on cold gas clouds in galaxy winds. Only a few works \citep{Farber2022,Chen2023,Richie2024arXiv} reach sufficiently low temperatures ($\sim10^2-10^3$~K) to include dust physics, while the ability of JWST to identify these structures motivates a need for more simulations of dust and molecular gas in outflows. 

The 10~pc scale clouds we observe in M\,82 reside inside the larger structures that we call plumes. In Fig.~\ref{fig:xray} we show that these plumes are generally associated to regions of lower X-ray brightness. This shows that the scenario of a hot wind surrounded by cool gas \citep{bolatto2013,Leroy2015} establishes itself very quickly, within 100 to 200~pc from the disk midplane, essentially at the point where the wind breaks out of the disk. 
The colocation of the PAH peaks and the CO peaks suggest that in the plume environment the local gas temperature is lower, thus allowing cold clouds to more easily survive.   \cite{Fielding2022} put forward a semianalytic theory where the large scale energetics of the wind are regulated, in part, by a turbulent mixing between cold and hot gas at the surface of clouds. The cloud sizes they consider are larger than our small scale clouds, and more similar to the plumes. 

The observations we present here are a starting point, while many questions still remain. 
How common is the substructure we identify in M\,82? Are the plumes and clouds common features of galactic winds? Is this substructure only a feature of starbursts, or does it extend to lower speed galactic fountains in main-sequence disks, where cosmic rays may be more important? Large scale galaxy evolution simulations find that galactic winds are the dominant mechanism to regulate galaxy growth \citep{Pillepich2018,Naab2017}. Observations, however, suggest that very little of the outflow gas fully escapes the halo \citep{Chisholm2015,Heckman2015,Davies2019,Marasco2023}. The energetics of the outflow are crucial to regulating star formation, and current theory suggest the small-scale structure of the cold gas is vital to setting those energetics \citep{Fielding2022}. JWST opens a window to the small-scale building blocks of galactic outflows --- more observations of PAH emission in such objects will allow us to build better models of galaxy evolution.

\section*{Acknowledgements}

This work is based on observations made with the NASA/ESA/CSA James Webb Space Telescope. The data were obtained from the Mikulski Archive for Space Telescopes at the Space Telescope Science Institute, which is operated by the Association of Universities for Research in Astronomy, Inc., under NASA contract NAS 5-03127 for JWST. These observations are associated with program JWST-GO-01701. Support for program JWST-GO-01701 is provided by NASA through a grant from the Space Telescope Science Institute, which is operated by the Association of Universities for Research in Astronomy, Inc., under NASA contract NAS 5-03127. 

D.B.F. acknowledges support by the Australian Research Council Centre of Excellence for All Sky Astrophysics in 3 Dimensions (ASTRO 3D), through project number CE170100013 and from Australian Research Council  Future Fellowship FT170100376

A.D.B. and S.A.C. acknowledge support from the NSF under award AST-2108140.

R.C.L. acknowledges support for this work provided by a NSF Astronomy and Astrophysics Postdoctoral Fellowship under award AST-2102625. 

L.A.L. and S. L. acknowledge support from the Heising-Simons Foundation grant 2022-3533 and NASA's Astrophysics Data Analysis Program under grant number 80NSSC22K0496. This work was also supported by NASA through Chandra Award Number AR4-25007X issued by the Chandra X-ray Center, which is operated by the Smithsonian Astrophysical Observatory for and on behalf of the National Aeronautics Space Administration under contract NAS8-03060.

I.D.L. acknowledges funding support from the European Research Council (ERC) under the European Union’s Horizon 2020 research and innovation programme DustOrigin (ERC-2019-StG-851622) and funding support from the Belgian Science Policy Office (BELSPO) through the PRODEX project “JWST/MIRI Science exploitation” (C4000142239). 

R.S.K. and S.C.O.G. acknowledge funding from the European Research Council via the Synergy Grant ``ECOGAL'' (project ID 855130), from the German Excellence Strategy via the Heidelberg Cluster of Excellence (EXC 2181 - 390900948) ``STRUCTURES'', and from the German Ministry for Economic Affairs and Climate Action in project ``MAINN'' (funding ID 50OO2206). 

V. V. acknowledges support from the scholarship ANID-FULBRIGHT BIO 2016 - 56160020, funding from NRAO Student Observing Support (SOS) - SOSPADA-015, and funding from the ALMA-ANID Postdoctoral Fellowship under the award ASTRO21-0062. 

L.L. acknowledges that a portion of their research was carried out at the Jet Propulsion Laboratory, California Institute of Technology, under a contract with the National Aeronautics and Space Administration (80NM0018D0004).

R.H.C. thanks the Max Planck Society for support under the Partner Group project "The Baryon Cycle in Galaxies" between the Max Planck for Extraterrestrial Physics and the Universidad de Concepción. R.H-C. also gratefully acknowledge financial support from ANID BASAL project FB210003.

\section*{Data Availability}

JWST data used in this will be available in the JWST MAST archive. The HST images are available in HST MAST archive. All other data is published.



\bibliographystyle{mnras}





\bsp	
\label{lastpage}
\end{document}

%% file: affiliations.tex
\newcommand{\AAPF}{NSF Astronomy and Astrophysics Postdoctoral Fellow}

\newcommand{\Arizona}{Steward Observatory, University of Arizona, Tucson, AZ 85721, USA}

\newcommand{\Maryland}{Department of Astronomy, University of Maryland, College Park, MD 20742, USA}

\newcommand{\JSI}{Joint Space-Science Institute, University of Maryland, College Park, MD 20742, USA}

\newcommand{\OSU}{Department of Astronomy, The Ohio State University, Columbus, OH 43210, USA}

\newcommand{\OSUCCAPP}{Center for Cosmology and Astro-Particle Physics, The Ohio State University, Columbus, OH 43210, USA}

\newcommand{\NRAOSocorro}{National Radio Astronomy Observatory, P.O. Box O, 1003 Lopezville Road, Socorro, NM 87801, USA}

\newcommand{\Kansas}{Department of Physics and Astronomy, University of Kansas, 1251 Wescoe Hall Dr., Lawrence, KS 66045, USA}

\newcommand{\MPIA}{Max-Planck-Institut f\"ur Astronomie, K\"onigstuhl 17, 69120 Heidelberg, Germany}

\newcommand{\IPAC}{IPAC, California Institute of Technology, 1200 East California Boulevard, Pasadena, CA 91125, USA}

\newcommand{\STScI}{Space Telescope Science Institute, 3700 San Martin Drive, Baltimore, MD 21218, USA}

\newcommand{\Swinburne}{Centre for Astrophysics and Supercomputing, Swinburne University of Technology, Hawthorn, VIC 3122, Australia}

\newcommand{\ASTROTD}{ARC Centre of Excellence for All Sky Astrophysics in 3 Dimensions (ASTRO 3D)}

\newcommand{\ITA}{Universit\"{a}t Heidelberg, Zentrum f\"{u}r Astronomie, Institut f\"{u}r Theoretische Astrophysik, Albert-Ueberle-Str. 2, D-69120 Heidelberg, Germany}

\newcommand{\IWR}{Universit\"{a}t Heidelberg, Interdisziplin\"{a}res Zentrum f\"{u}r Wissenschaftliches Rechnen, Im Neuenheimer Feld 205, D-69120 Heidelberg, Germany}

\newcommand{\UTA}{Department of Astronomy, The University of Texas at Austin, 2515 Speedway, Stop C1400, Austin, TX 78712, USA}

\newcommand{\UWYO}
{Department of Physics \& Astronomy, University of Wyoming, Laramie, WY 82071, USA}

\newcommand{\JPL}
{Jet Propulsion Laboratory, California Institute of Technology, 4800 Oak Grove Dr., Pasadena, CA 91109, USA}

\newcommand{\Leiden}
{Leiden Observatory, Leiden University, P.O.~Box 9513, 2300~RA~Leiden, The Netherlands}

\newcommand{\UdeC}{Departamento de Astronom\'ia, Universidad de Concepci\'on, Barrio Universitario, Concepci\'on, Chile}

\newcommand{\ESOST}{European Space Agency, c/o STScI, 3700 San Martin Drive, Baltimore, MD 21218, USA}

\newcommand{\INAOE}{Instituto Nacional de Astrof\'{\i}sica, \'Optica y Electr\'onica, Luis Enrique Erro 1, Tonantzintla 72840, Puebla, Mexico}

\newcommand{\NMMT}{New Mexico Institute of Mining and Technology, 801 Leroy Place, Socorro, NM 87801, USA}

\newcommand{\UToledo}{Ritter Astrophysical Research Center, University of Toledo, Toledo, OH 43606, USA}

\newcommand{\Cornell}{Department of Astronomy, Cornell University, Ithaca, NY 14853, USA}

\newcommand{\Flatiron}{Center for Computational Astrophysics, Flatiron Institute, 162 5th Ave, New York, NY 10010, USA}

\newcommand{\UGR}{Dept. Física Teórica y del Cosmos, Universidad de Granada, Granada, Spain}

\newcommand{\UGent}{Sterrenkundig Observatorium, Ghent University, Krijgslaan 281 - S9, B-9000 Gent, Belgium}

\newcommand{\France}{Coll\`ege de France, 11 Pl. Marcelin Berthelot, 75231 Paris, France}

\newcommand{\ParisObs}{Observatoire de Paris, 61 avenue de l’Observatoire, 75014 Paris, France}